\documentclass{iopart}

\jl{6}        % to appear in journal: Classical and Quantum Gravity
\eqnobysec    % section equation numbering

\def\beq{\begin{equation}}
\def\eeq{\end{equation}}

\def\dualp#1{{}^{\ast_{(\hbox{$\scriptstyle #1$})}} \kern-1pt}
\def\dual{{}^{\ast} \kern-1pt}

\def\rmd{{\rm d}}   

\begin{document}

\title[Circular holonomy  in the Taub-NUT spacetime]
{Circular holonomy in the Taub-NUT spacetime}

\author{
Donato Bini\dag ${}^\ast$,
Christian Cherubini\ddag ${}^\ast$, 
Robert T. Jantzen$\S{}^\ast$
}

\address{
  \dag\
  Istituto per le Applicazioni del Calcolo ``M. Picone", C.N.R.,
   I-- 00161 Roma, Italy \\
}
\address{
  ${}^\ast$
ICRA International Center for Relativistic Astrophysics, University of Rome, I--00185 Rome, Italy
}

\address{
  \S\
  Department of Mathematical Sciences, Villanova University, 
  Villanova, PA 19085, USA
}
\address{
  \ddag\
Dipartimento di Fisica ``E.R.
Caianiello'', Universit\`a di Salerno, I--84081, Italy
}

\begin{abstract}
Parallel transport around closed circular orbits in the equatorial plane of the Taub-NUT spacetime is analyzed to reveal the effect of the gravitomagnetic monopole parameter on circular holonomy transformations.
Investigating the boost/rotation decomposition of the connection 1-form matrix evaluated along these orbits, one finds a situation that reflects the behavior of the general orthogonally transitive stationary axisymmetric case and indeed along Killing trajectories in general.
\end{abstract}

\pacs{0420C}

\submitted  June 18, 2002

\section{Introduction}

Recently Rothman, Ellis and Murugan~\cite{rot} gave a detailed analysis of parallel transport of vectors along circles in the Schwarzschild spacetime. This work was then generalized to circles in the equatorial plane of the Kerr spacetime by Maartens, Mashhoon and Matravers~\cite{MMM} and finally to helices (arbitrary stationary circular orbits) in the equatorial plane of the Kerr spacetime by Bini, Jantzen and Mashhoon~\cite{bjm}.
The last investigation also examined the relationship between holonomy invariance and clock effects and began studying how the presence of a gravitomagnetic monopole term in the Kerr-Taub-NUT spacetime \cite{mistau,Miller,Carter,DemNew,zimsha,lynnou,noulyn}
changes the clock effect discussion from that of the Kerr spacetime.
Although this spacetime is not asymptotically flat, like the G\"odel spacetime it can play a useful role in helping understand how gravitomagnetism works.

Here we  discuss how circular holonomy invariance is modified for closed circular trajectories in the equatorial plane of the simpler Taub-NUT spacetime, a special case of the Kerr-Taub-NUT family which nicely illustrates the only really new feature which the NUT parameter introduces.
The situation is completely different from the Schwarzschild and Kerr cases since the NUT parameter induces an additional boost in a plane orthogonal to the plane of the parallel transport rotation already present in those spacetimes (called a 4-screw Lorentz transformation by Synge \cite{synge}). 

This is in fact typical of parallel transport along general closed circular orbits in any stationary axisymmetric spacetime, including those off the equatorial plane in Kerr 
\cite{bogiati}, which can be analyzed in a similar way through the decomposition of the induced connection matrix along these Killing trajectories into its electric and magnetic parts when interpreted as a 2-form after lowering its contravariant index.
Turning on the rotation in the Taub-NUT spacetime or in its special case, the Schwarzschild spacetime, to yield the Kerr-Taub-NUT or Kerr spacetime respectively essentially makes the associated Poynting vector nonzero and aligned with the axial Killing vector direction, so that a boost in this direction can be chosen to align the electric and magnetic parts along a common direction.

Because both parallel transport and Lie dragging along a Killing trajectory preserve inner products, a pair of orthonormal frames which undergo these two respective transports along such a trajectory are related by a parameter-dependent Lorentz transformation. If one works with a stationary axisymmetric frame, whether orthonormal or not, 
the induced connection matrix evaluated on the tangent vector to a Killing trajectory always defines an associated stationary axisymmetric 2-form (in index-lowered form) which corresponds to the Lie algebra generator (in mixed tensor form) of the parameter-dependent Lorentz transformation. 
This mixed tensor can be classified by its eigenvalue/eigenspace properties (see appendix A), thus geometrically characterizing the parameter-dependent Lorentz transformation between parallel transport and Lie dragging along the Killing trajectory which results from its exponentiation. 

The Taub-NUT spacetime provides a simple example which illustrates the generic case with four distinct (nonzero) eigenvalues, and is therefore typical of the most general situation which can occur along Killing trajectories in any spacetime with symmetry. 
This immediately explains why there are no circular geodesics in the Taub-NUT equatorial plane, since this requires the Lorentz transformation generator to have the zero eigenvalue to allow an invariantly parallel transported circular orbit tangent vector. In fact the locus of radii off the equatorial plane where the circular geodesics do occur is exactly the pair of surfaces symmetric about the equatorial plane where this Lorentz generator admits a zero eigenvalue, namely where its electric and magnetic parts are orthogonal to each other.

Holonomy invariance can be considered for closed circular orbits in the Taub-NUT spacetime, but only for those special tangent vectors which lie in the plane in which the parallel transport rotation takes place. One finds a situation similar to the Schwarzschild case in which a band of holonomy invariance exists from the horizon out to infinity.

\section{The Taub-NUT metric}

The metric of the Taub-NUT spacetime~\cite{mistau,Miller,Carter,DemNew} in Boyer-Lindquist-like coordinates is
\beq\fl\qquad
\rmd s^2
=-\Sigma^{-1}\Delta(\rmd t + 2\ell \cos\theta \rmd\phi)^2
   +\Sigma(\Delta^{-1} \rmd r^2 +\rmd \theta^2 +\sin^2\theta  \rmd\phi^2)
\eeq
where
$\Sigma = r^2 + \ell^2$,
$\Delta = r^2-2Mr-\ell^2$.
The parameters $M$ and $\ell$ are associated with the mass and the gravitational magnetic monopole strength of the source. The metric is reflection-symmetric ($\theta\to\pi-\theta$) about the equatorial plane $\theta=\pi/2$ only when $\ell=0$.

It is convenient to introduce the usual zero-angular-momentum observer (ZAMO) orthonormal frame $e_{\hat\alpha}$, $\alpha = 0,r,\theta,\phi$
which normalizes the spatial coordinate frame vectors
$e_{\hat r}=g_{rr}^{-1/2}\partial_r,
 e_{\hat\theta}=g_{\theta\theta}^{-1/2}\partial_\theta,
 e_{\hat\phi}=g_{\phi\phi}^{-1/2}\partial_\phi$ 
and completes the frame with the unit normal to the $t$-coordinate hypersurfaces
$e_{\hat 0} = n = N^{-1}(\partial_t - N^\phi\partial_\phi)$, where $N=(-g^{tt})^{-1/2}$ is the lapse function and
$N^\phi=g_{t\phi}/g_{\phi\phi}$ the only nonzero component of the shift vector. $e_{\hat 0}$ is interpreted as the 4-velocity of the ZAMOs. 
The ZAMO dual 1-forms are
\begin{eqnarray}
\omega^{\hat 0}&=& [\Delta/g_{\phi\phi}]^{1/2} \sin\theta \, \rmd t
  \ ,\quad
\omega^{\hat r}= [(r^2+l^2)/\Delta]^{1/2}\rmd r
  \ ,\nonumber \\
\omega^{\hat \theta}&=&(r^2+l^2)^{1/2}\rmd \theta 
  \ ,\quad
\omega^{\hat \phi}= g_{\phi\phi}^{1/2} 
        [ \rmd \phi -\frac{2\ell\Delta \cos\theta}{g_{\phi\phi}(r^2+l^2)}\rmd t]
\ ,
\end{eqnarray}
with
\beq
g_{\phi\phi}=(r^2+l^2) \sin^2\theta  - \frac{4l^2\Delta \cos^2\theta}{r^2+l^2}
\ .
\eeq
This allows one to identify the lapse and shift
\beq
N= [\Delta/g_{\phi\phi}]^{1/2}  \sin\theta 
\ ,\qquad
N^\phi= -\frac{2\ell\Delta \cos\theta}{g_{\phi\phi}(r^2+l^2)} \ .
\eeq
The 1-form $\omega^{\hat 0}$ (or vector field $n=e_{\hat 0}$) is timelike outside the horizon, which occurs at $\Delta=0$, namely for $r>r_+=M+(M^2+l^2)^{1/2}$.

\section{Parallel transport along $\phi$-circles in the Taub-NUT spacetime}

In the Taub-NUT spacetime consider the parallel transport of a vector 
$X = X^\alpha \partial_\alpha$ along a $\phi$-circle 
parametrized by the coordinate $\phi$, with unit tangent vector 
$e_{\hat \phi}$ on the equatorial plane $\theta=\pi/2$
\beq\label{eq:A}
\nabla_\phi X^\alpha =
\frac{\rmd X^\alpha}{\rmd \phi} -A^\alpha{}_\beta X^\beta =0\ ,
\eeq
where $A_{\alpha\beta} = - g_{\phi[\alpha,\beta]}$ define the components of a 2-form  \beq
  A^\flat = -\sigma_{(B)} \omega^{\hat 0}\wedge \omega^{\hat\theta}
     +\sigma_{(R)} \omega^{\hat r}\wedge \omega^{\hat\phi} 
        = -\omega^{\hat 0}\wedge \mathcal{E}_{\hat\theta} \omega^{\hat\theta}  + \dualp{n} [\mathcal{B}_{\hat\theta} \omega^{\hat\theta} ]
\eeq
which corresponds to the mixed tensor whose sign-reversed component matrix is just the coordinate connection 1-form 
$\omega^\alpha{}_\beta$ evaluated along the Killing vector $\partial_\phi$
\beq
 -A^\alpha{}_\beta = \Gamma^\alpha{}_{\phi\beta} =\omega^\alpha{}_\beta(\partial_\phi)\ .
\eeq
The notation $\dualp{n} \mathcal{B}$ stands for the duality operation in the 3-space orthogonal to the observer 4-velocity $n$, for example
$\dualp{n}\omega^{\hat\theta} = -\omega^{\hat r}\wedge\omega^{\hat\phi}$.

The index-lowered antisymmetry $A_{(\alpha\beta)}=0$
means that $A$ generates a Lorentz transformation of the tangent space as a mixed tensor acting on vectors by right contraction. Its coordinate components are constant along the curve, corresponding to the fact that the corresponding tensor is Lie dragged along this Killing trajectory. Its component matrix differs from an element of the Lie algebra of the Lorentz group only by a constant linear transformation, corresponding to re-expressing it in a stationary axisymmetric orthonormal frame. 
In fact, since any two stationary axisymmetric frames along these Killing trajectories are related to each other by a constant linear transformation, one can express the parallel transport equation and its solution in any such frame using this constant linear transformation freedom. This makes it easy to interpret directly the Lorentz transformation which is generated by the tensor 
$A^\alpha{}_\beta e_\alpha \otimes \omega^\beta
=A^{\hat\alpha}{}_{\hat\beta} e_{\hat\alpha}\otimes \omega^{\hat\beta}$ by examining its action in the closely related ZAMO orthonormal spherical frame.

The electric and magnetic parts of the 2-form $A^\flat$ are aligned along $e_{\hat\theta}$ and are then explicitly (outside the horizon where $\Delta>0$)
\begin{eqnarray}
\fl\qquad
\mathcal{E}_{\hat\theta} = \sigma_{(B)}
&=& %- correction in proof
\frac12 
g_{\phi\phi} \left(\frac{-g^{tt}}{g_{\theta\theta}} \right)^{1/2} \left(\frac{g_{t\phi}}{g_{\phi\phi}}\right)_{,\theta} 
= -% correction in proof
g_{\phi\phi}^{1/2}\theta_{\hat\phi\hat\theta }
         = \frac{l\Delta^{1/2}}{r^2+l^2}   \ ,\nonumber\\
\fl\qquad
\mathcal{B}_{\hat\theta} = \sigma_{(R)}
&=& \frac12 \frac{1}{(g_{\phi\phi}g_{rr})^{1/2}}g_{\phi\phi,r}
        = g^{1/2}_{\phi\phi,\hat r}=-g_{\phi\phi}^{1/2}\, k_{\rm(lie)}{}_{\hat r }
        = \frac{r\Delta^{1/2}}{r^2+l^2}
  \ ,
\end{eqnarray}
with
$\mathcal{E}_{\hat\theta}/\mathcal{B}_{\hat\theta} =\ell/r$.
Here the Lie curvature vector (1-form) \cite{idcf1,circfs} of the circle $k_{\rm(lie)} = -\rmd\ln g_{\phi\phi}^{1/2}$ 
associated with the intrinsic curvature of the equatorial plane in the constant $t$ hypersurface containing the circle induces the magnetic part of this Lorentz generator (a space curvature effect), while the expansion tensor $\theta(n)_{\alpha\beta} =-K_{\alpha\beta}$ of the unit normal vector field $n$, equivalently the sign-reversed extrinsic curvature tensor of this hypersurface, induces the  electric part of the Lorentz generator. It is this expansion tensor (symmetric part of the gravitomagnetic tensor \cite{mfg}) which is associated with gravitomagnetic effects. Since the closed $\phi$ circles lie in constant $t$ hypersurfaces, gravitoelectric effects are not directly involved.

The electric part of the 2-form generates a boost in the $e_{\hat0}$-$e_{\hat\theta}$ plane, while the magnetic part generates a rotation in the $e_{\hat r}$-$e_{\hat\phi}$ plane.
Since the electric and magnetic parts of $A$ are aligned along the same direction, these two planes are orthogonal and each is separately invariant under the action of $A$ as a linear transformation.
One finds $X(\phi) = e^{\phi A} X(0)$, which when re-expressed in the orthonormal frame is equivalent to the Lorentz 4-screw
\begin{eqnarray}
\left( \begin{array}{c} 
X^{\hat 0}(\phi)\\ 
X^{\hat \theta}(\phi)\\
\end{array}\right)
&=& 
\left( \begin{array}{cc}
\cosh \sigma_{(B)}\phi & 
\sinh \sigma_{(B)}\phi \\
\sinh \sigma_{(B)}\phi & 
\cosh \sigma_{(B)}\phi \\
\end{array}\right) 
\left( \begin{array}{c}
X^{\hat 0}(0) \\
X^{\hat \theta}(0) \\
\end{array}\right)
   \ ,\nonumber \\
\left( \begin{array}{c} 
X^{\hat r}(\phi)\\ 
X^{\hat \phi}(\phi)\\
\end{array}\right)
&=& 
\left( \begin{array}{cc}
\phantom{-}\cos \sigma_{(R)} \phi & 
\sin \sigma_{(R)}\phi \\
-\sin\sigma_{(R)} \phi & 
\cos\sigma_{(R)}\phi \\
\end{array}\right) 
\left( \begin{array}{c}
X^{\hat r}(0) \\
X^{\hat \phi}(0) \\
\end{array}\right)
  \ .
\end{eqnarray} 
Since the rapidity (namely the hyperbolic angle defining a boost) coefficient $\sigma_{(B)}$ satisfies $\sigma_{(B)} \propto l $, it vanishes in the 
Schwarzschild limit $l=0$, corresponding to an invariant 2-plane under parallel transport, 
while the parallel transport frequency satisfies $\sigma_{(R)}-1 \neq0$ in that limit only if $M\neq0$.

The discussion of parallel transport is now quite different from the Kerr spacetime equatorial plane case because of the presence of an electric part of the matrix $A$.
Here the rapidity coefficient $\sigma_{(B)}$ is nonzero even on the equatorial plane because of the reflection symmetry breaking.
Moreover, the transported vector undergoes a simultaneous rotation and boost in fixed orthogonal 2-planes in contrast with the Kerr equatorial plane case where only a rotation in a fixed 2-plane is present (sufficiently far from the black hole).

Holonomy invariance can clearly only occur for vectors orthogonal to the boost plane.
By restricting the discussion to vectors with only $r$ and $\phi$ components,
holonomy invariance occurs when the following condition is satisfied
\beq
\sigma_{(R)} = \frac{r\Delta^{1/2}}{r^2+l^2}= p/n\ ,
\label{holo}
\eeq
where $n\neq0$ and $p$ are two integers of the same sign (if $p\neq0$), corresponding to $p$ parallel transport loops occurring during $n$ $\phi$-coordinate orbital loops. $p>0$ corresponds to a clockwise rotation in the $e_{\hat r}$-$e_{\hat\phi}$ plane, while $n>0$ corresponds to a counterclockwise orbital rotation.
Since $0\leq \sigma_{(R)}<1$ for all $r$ outside the horizon where $\Delta=0=\sigma_{(R)}$ and since $\sigma_{(R)}\to1$ as $r\to\infty$, this condition is satisfied for each pair of integers for which $p/n<1$. Thus for these special vectors tangent to the equatorial plane, one finds a band of holonomy invariance extending out from the horizon to infinity
exactly as in the Schwarzschild case.
The case $l=0$ reduces to the Schwarzschild result for the relation between radius and loop numbers
\beq
r=\frac{2M}{1-p^2/n^2}\ .
\eeq 

\section{Orthogonally transitive stationary axisymmetric spacetimes}

The Kerr-Taub-NUT spacetime family containing the Taub-NUT metric itself belongs to the larger family of orthogonally transitive stationary axisymmetric vacuum spacetimes \cite{kramer}. Apart from a boost of the pair of orthogonal 2-planes of the Lorentz 4-screw, the picture for Taub-NUT parallel transport around a circle in the equatorial plane essentially holds for any generic closed circular paths in this larger family as well. 
In Boyer-Lindquist-like coordinates the more general metric is
\beq
\rmd s^2 
 =  g_{tt}\rmd t^2 +2g_{t\phi}\rmd t \rmd \phi +g_{\phi\phi}\rmd \phi^2 
  + g_{rr}\rmd r^2 + g_{\theta\theta}\rmd \theta^2\ ,
\eeq
while the ZAMO adapted dual frame is
\beq
\fl\qquad
\omega^{\hat 0} = N \rmd t \ ,\
\omega^{\hat r} = g_{rr}^{1/2} \rmd r \ ,\
\omega^{\hat \theta} = g_{\theta\theta}^{1/2} \rmd \theta \ ,\
\omega^{\hat \phi}= g_{\phi\phi}^{1/2}[\rmd \phi + \frac{g_{t\phi}}{g_{\phi\phi}} \rmd t ]\ .
\eeq

The parallel transport equation (\ref{eq:A}) has the same form but now
the value of the coordinate connection 1-form matrix along $\partial_\phi$ is instead 
\begin{eqnarray}\label{Aktheta}
\fl\qquad
-A^\flat &=& 
 \frac12 g_{t\phi,r} \rmd t\wedge \rmd r 
+\frac12 g_{t\phi,\theta} \rmd t\wedge \rmd \theta  
+\frac12 g_{\phi\phi,r} \rmd \phi\wedge \rmd r  
+\frac12 g_{\phi\phi,\theta} \rmd \phi\wedge \rmd \theta   
\ ,\nonumber\\ \fl\qquad
  &=&\frac12 \rmd t\wedge \rmd  g_{t\phi} + \frac12 \rmd \phi\wedge \rmd  g_{\phi\phi} 
%\ . correction in proof
\ ,\nonumber\\ \fl\qquad
  &=& \omega^{\hat 0}\wedge \mathcal{E}  
        - \dualp{n} [\mathcal{B} ]
\ ,
\end{eqnarray}
where $\dualp{n} X$ and $X\times_n Y = \dualp{n} [X\wedge Y]$ are the spatial dual and cross product of spatial 1-forms in the geometry of the constant $t$ hypersurface. 
The electric and magnetic parts of $A^\flat$ are respectively
\beq
  \mathcal{E} =  
-%correction in proof
g^{1/2}_{\phi\phi} \theta_{\hat \phi} 
\ ,\quad
  \dualp{n} [\mathcal{B} ] 
   = g^{1/2}_{\phi\phi} \omega^{\hat \phi}\wedge k_{\rm(lie)}\ ,
\eeq
where the 1-form $\theta_{\hat \phi} $ is the contraction of the shear tensor $\theta(n)_{\alpha\beta}$ with $e_{\hat\phi}$ 
and $k_{\rm(lie)}$
is the Lie intrinsic curvature vector (1-form) of the $\phi$ coordinate circles,
explicitly
\begin{eqnarray}
\label{thetaphi}
\theta_{\hat \phi} 
&=& \theta_{\hat \phi\hat r}\omega^{\hat r}
   +\theta_{\hat \phi\hat \theta}\omega^{\hat \theta} 
    \nonumber\\
&=& -\frac{g^{1/2}_{\phi\phi}}{2N} 
  \left[\left(\frac{g_{t\phi}}{g_{\phi\phi}}\right)_{,\hat r}\omega^{\hat r}
   +\left(\frac{g_{t\phi}}{g_{\phi\phi}}\right)_{,\hat \theta}\omega^{\hat \theta}\right]
  = - \frac{g^{1/2}_{\phi\phi}}{2N} \rmd \left(\frac{g_{t\phi}}{g_{\phi\phi}}\right)
  \ ,\nonumber \\
k_{\rm(lie)} &=& k_{\rm(lie)}{}_{\hat r}\omega^{\hat r}+ k_{\rm(lie)}{}_{\hat \theta}\omega^{\hat \theta}
  \nonumber \\
& =& -[(\ln g^{1/2}_{\phi\phi})_{,\hat r}\omega^{\hat r}
+(\ln g^{1/2}_{\phi\phi})_{,\hat \theta}\omega^{\hat \theta}]
= - \rmd \ln g^{1/2}_{\phi\phi}\ .
\end{eqnarray}
The magnetic part 1-form and the Poynting vector cross product (1-form) are then explicitly
\beq\fl\quad
   \mathcal{B} 
%=k_{\rm(lie)}\times_n e_{\hat \phi}= -k_{\rm(lie)}{}_{\hat\theta}\omega^{\hat r}+
%k_{\rm(lie)}{}_{\hat r}\omega^{\hat \theta}\ ,\quad
%    \mathcal{E} \times_n \mathcal{B} 
%         = \theta_{\hat \phi} \cdot k_{\rm(lie)} \, \omega^{\hat\phi}\ ,
= g_{\phi\phi}^{1/2} e_{\hat \phi}\times_n k_{\rm(lie)}
= g_{\phi\phi}^{1/2}[-k_{\rm(lie)}{}_{\hat\theta}\omega^{\hat r}+
k_{\rm(lie)}{}_{\hat r}\omega^{\hat \theta}]\ ,\quad
    \mathcal{E} \times_n \mathcal{B} 
         = -g_{\phi\phi} [\theta_{\hat \phi} \cdot k_{\rm(lie)}] \omega^{\hat\phi}\ ,
\eeq
while the scalar invariants of $A$ (see appendix A) are
\begin{eqnarray}
   \mathcal{E} \cdot \mathcal{B} 
&=& g_{\phi\phi} [k_{\rm(lie)}{}_{\hat r}\theta_{\hat \phi\hat \theta}
                 -k_{\rm(lie)}{}_{\hat \theta}\theta_{\hat \phi\hat r}]
= g_{\phi\phi}[\theta_{\hat \phi} \times_n k_{\rm(lie)}]_{\hat \phi} \ , \nonumber\\
\mathcal{E}^2 -\mathcal{B}^2 
&=& g_{\phi\phi}[||\theta_{\hat \phi}||^2-||k_{\rm(lie)}||^2] \ .
\end{eqnarray} 
Note that $\mathcal{E} \cdot \mathcal{B}\neq0$ unless $k_{\rm(lie)}$  and $\theta_{\hat \phi}$ are parallel.

One can recover the simpler picture of the Taub-NUT case by simply performing a boost in the $e_{\hat\phi}$ direction to align the electric and magnetic parts of $A$ as described in exercise 20.6 of Misner, Thorne and Wheeler \cite{mtw}. 
A boost along $\phi$ has the form
\begin{eqnarray}\label{eq:boost}
n&=& \cosh \alpha\, U +\sinh \alpha\, E_{\hat \phi}\ , \quad
e_{\hat \phi}  = \sinh \alpha\, U + \cosh \alpha\, E_{\hat \phi} \ ,
\end{eqnarray}
where $U$, $E_{\hat \phi}$ is an orthogonal pair of respectively timelike and spacelike unit vectors in the $e_{\hat 0}$-$e_{\hat\phi}$ plane. 
Re-expressing $A$ in terms of these new vectors, one has
\begin{eqnarray}
\fl\quad
 g_{\phi\phi}^{-1/2} A^\flat 
        &=& -[\cosh \alpha\, U +\sinh \alpha\, E_{\hat \phi}^\flat ]\wedge \theta_{\hat \phi}  
        +[\sinh \alpha\, U + \cosh \alpha\, E_{\hat \phi}^\flat ] \wedge k_{\rm(lie)} 
\ ,\nonumber\\ \fl\quad
  &=& U\wedge [- \cosh \alpha\, \theta_{\hat \phi} +\sinh \alpha\, k_{\rm(lie)}]
+E_{\hat \phi}^\flat \wedge [-\sinh \alpha\, \theta_{\hat \phi} + \cosh \alpha\, k_{\rm(lie)} ]
\nonumber\\ \fl\quad
&\equiv & g_{\phi\phi}^{-1/2}[ \sigma_{(B)}(U) U \wedge E_{\hat \theta}^\flat + 
\sigma_{(R)}(U)E_{\hat \phi}^\flat \wedge E_{\hat r}^\flat ]
\ ,
\end{eqnarray}
\typeout {[The order of vector $E_r$, $E_\theta$, $E_\phi$ is correct: this is a long calculation! ]}
\\
which defines two additional unit vectors $E_{\hat r}$ and $E_{\hat \theta}$ lying in the $r$-$\theta$ plane which in turn determine the new electric and magnetic parts of $A^\flat$ with respect to $U$, namely $\sigma_{(B)}(U) E_{\hat \theta}$ and $\sigma_{(R)}(U) E_{\hat \phi}^\flat \times E_{\hat r}^\flat$. 
Requiring that $E_{\hat r}$ be orthogonal to $E_{\hat \theta}$, i.e. $E_{\hat r}\cdot E_{\hat \theta} =0$, is the condition which aligns the electric and magnetic part 1-forms along $E_{\hat \theta}^\flat $ in this context and makes $\{U,E_{\hat r},E_{\hat \theta},E_{\hat\phi}\}$ an orthonormal frame
\beq\label{eq:tanh}
\tanh 2 \alpha 
= \frac{2\theta_{\hat \phi} \cdot k_{\rm(lie)} }{||\theta_{\hat \phi} ||^2+ ||k_{\rm(lie)}||^2}\rightarrow
\alpha
=\frac12 \log \left(\frac{||\theta_{\hat \phi}+ k_{\rm(lie)}||}{ ||\theta_{\hat \phi}- k_{\rm(lie)}|| }\right)\ ,
\eeq
where the identity $\tanh^{-1}x = \frac12 \ln\frac{1+x}{1-x}$ has been used together with some minor dot product algebra. 
Note that since the $e_{\hat r}$-$e_{\hat\theta}$ plane is always spacelike, $|\tanh 2 \alpha| \leq1$%correction in proof
, while for comparison with the formula of Misner, Thorne and Wheeler, an equivalent form of this result is
\beq
\tanh 2\alpha = -\frac{2(\mathcal{E} \times_n \mathcal{B})\cdot e_{\hat \phi}}{\mathcal{E}^2 +\mathcal{B}^2}
\quad{\rm or}\quad
|\tanh 2\alpha| = \frac{2||\mathcal{E} \times_n \mathcal{B}||}{\mathcal{E}^2 +\mathcal{B}^2}
\ .
\eeq
One then finds
\beq
\fl
\sinh \alpha = \frac12 \frac{||\theta_{\hat \phi}+ k_{\rm(lie)}||- ||\theta_{\hat \phi}- k_{\rm(lie)}||}{\sqrt{||\theta_{\hat \phi}+ k_{\rm(lie)}||\,||\theta_{\hat \phi}- k_{\rm(lie)}||}},\,
\cosh \alpha = \frac12 \frac{||\theta_{\hat \phi}+ k_{\rm(lie)}||+ ||\theta_{\hat \phi}- k_{\rm(lie)}||}{\sqrt{||\theta_{\hat \phi}+ k_{\rm(lie)}||\,||\theta_{\hat \phi}- k_{\rm(lie)}||}}\ .
\eeq

In the region of spacetime where $|\tanh 2\alpha| < 1$, 
one has exactly the same situation as in the Taub-NUT equatorial plane case
in the new orthonormal frame $\{U,E_{\hat r},E_{\hat \theta},E_{\hat\phi}\}$.
Two of the components of the transported vector 
undergo a boost in the $U$-$E_{\hat \theta}$ plane with rapidity coefficient 
\begin{eqnarray}
\label{sigmaB}
\fl
\sigma_{(B)}
&=&g^{1/2}_{\phi\phi} \left(\frac12 [||\theta_{\hat \phi}+k_{\rm(lie)} ||
\, ||\theta_{\hat \phi}-k_{\rm(lie)} ||
         + ||\theta_{\hat \phi} ||^2 - ||k_{\rm(lie)}||^2
]\right)^{1/2}
\nonumber\\\fl
&=&g^{1/2}_{\phi\phi} \left(\frac12 [\sqrt{(||\theta_{\hat \phi}||^2+||k_{\rm(lie)}||^2)^2-4(\theta_{\hat \phi}\cdot k_{\rm(lie)})^2 }
         + ||\theta_{\hat \phi} ||^2
- ||k_{\rm(lie)}||^2
]\right)^{1/2}
,
\end{eqnarray}
and the other two undergo a rotation in the orthogonal $E_{\hat r}$-$E_{\hat \phi}$ plane with frequency 
\begin{eqnarray}\label{sigmaR}
\fl\quad
\sigma_{(R)}
&=&g^{1/2}_{\phi\phi} \left(\frac12 \left[||\theta_{\hat \phi}+k_{\rm(lie)} ||
\, ||\theta_{\hat \phi}-k_{\rm(lie)} || 
         - ||\theta_{\hat \phi} ||^2+ ||k_{\rm(lie)}||^2
\right]\right)^{1/2}
\nonumber\\\fl\quad
&=&g^{1/2}_{\phi\phi} \left(\frac12 \left[\sqrt{(||\theta_{\hat \phi}||^2+||k_{\rm(lie)}||^2)^2-4(\theta_{\hat \phi}\cdot k_{\rm(lie)})^2 }
         - ||\theta_{\hat \phi} ||^2+ ||k_{\rm(lie)}||^2
\right]\right)^{1/2}
\end{eqnarray}
and  
the two invariants are just
\beq
\mathcal{E}\cdot \mathcal{B}= \sigma_{(B)}\sigma_{(R)} \ ,\
\mathcal{E}^2- \mathcal{B}^2 =  \sigma_{(B)}^2-\sigma_{(R)}^2 \ .
\eeq

The condition $\tanh 2 \alpha =\pm 1$  marks a transition surface where respectively  $\theta_{\hat \phi}=\pm k_{\rm(lie)}$,  causing both scalar invariants of $A$ to vanish and
$A$ becomes decomposable
\beq
A=g_{\phi\phi}^{1/2} [ \mp \omega^{\hat 0} +\omega^{\hat\phi} ]\wedge k_{\rm(lie)}\ .
\eeq
Since $\mp \omega^{\hat 0} +\omega^{\hat\phi}$ is null and $k_{\rm(lie)}$ is spacelike,
$A$ clearly generates a null rotation as it should (see appendix A).

In the Taub-NUT case $\theta_{\hat \phi}$ is orthogonal to $k_{\rm(lie)}$:
$\theta_{\hat \phi} \cdot k_{\rm(lie)} =0$, so $\alpha =0$ and
\beq
\sigma_{(R)}=g^{1/2}_{\phi \phi}||\theta_{\hat \phi}||\ ,\quad 
\sigma_{(B)}=g^{1/2}_{\phi\phi} ||k_{\rm(lie)}||\ .
\eeq
In the Kerr equatorial plane case studied in \cite{bjm},
$\theta_{\hat \phi}$ is instead parallel to $k_{\rm(lie)}$ and both of them lie along $e_{\hat r}$,
so that $(\theta_{\hat \phi}\cdot k_{\rm(lie)})^2=||\theta_{\hat \phi}||^2|| k_{\rm(lie)}||^2$.
The sign-reversed ratio of their radial components defines the geodesic meeting point observer ZAMO relative velocity  (see \cite{idcf2}, discussion after Eq.~(4.6), and \cite{bjm})
\beq
\label{gmp}
\nu_{\rm (gmp)} = - \theta_{\hat \phi \hat r}/k_{\rm(lie)}{}_{\hat r}\ .
\eeq
The expressions in Eqs.~(\ref{sigmaB}), (\ref{sigmaR}) then reduce to Heaviside functions in $\pm[||k_{\rm(lie)}||^2 - ||\theta_{\hat \phi} ||^2] = \pm\gamma_{\rm (gmp)}^{-2} ||k_{\rm(lie)}||^2$
\beq
 [\sigma_{(B)}, \sigma_{(R)}]
= g^{1/2}_{\phi \phi}||k_{\rm(lie)}|| \gamma_{\rm (gmp)}^{-1} \,  
[H (\nu_{\rm (gmp)}^2 -1),H (1-\nu_{\rm (gmp)}^2)]
\ ,
\eeq
where $\gamma_{\rm (gmp)}=|1-\nu_{\rm (gmp)}^2|^{-1/2}$.
Thus one only has a rotation outside the geodesic meeting point observer horizon at 
$|\nu_{\rm (gmp)}|=1$ and only a boost inside, corresponding to the two semi-singular cases of appendix A, with the transition surface corresponding to the singular null rotation case.

\section{Concluding Remarks}

Circular orbits play a fundamental role in the geometry of orthogonally transitive stationary axisymmetric spacetimes and the properties of parallel transport along them determine the behaviour of test particles and test gyroscopes among other physically interesting phenomena. Here some key features of this parallel transport have been eludicated for closed circular orbits in general and for the equatorial plane of the Taub-NUT spacetime in particular. The question of holonomy invariance is at once both similar to and very different from the corresponding Schwarzschild spacetime. No general holonomy invariance exists for the Taub-NUT case, but a restricted such invariance does exist for a subspace of the tangent space, with properties similar to those of Schwarzschild. 

For the general case of orthogonally transitive stationary axisymmetric symmetry,
one sees how the intrinsic and extrinsic curvature properties of the constant time slices in which these circular orbits lie directly translate into magnetic and electric parts of the Lorentz transformation generator governing parallel transport along them, associated with space curvature and gravitomagnetic effects respectively. There remains only the complication of considering the more general helical circular orbits along a general (fixed) Killing direction, where the gravitoelectric field (acceleration of the ZAMOs) contributes to a boost of the eigenspaces of the Lorentz generator governing parallel transport relative to the closed circular orbit case.  Studying the interesting case of timelike geodesics will then help clarify how the magnetic monopole of the Kerr-Taub-NUT family affects test particle motion and generalized clock and spin comparison effects.

\appendix

\section{Classification of Lorentz generators}

Suppose $\mathcal{A}$ is a linear transformation of the tangent space generating a Lorentz transformation $e^{\lambda \mathcal{A}}$. Then $\mathcal{A}$ identifiable with a $1\choose1$-tensor so that $\mathcal{A}^\flat$ is a 2-form which can be decomposed with respect to any timelike unit vector $u$ into its electric and magnetic parts as
\beq
\mathcal{A}^\flat = 
%- correction in proof
u^\flat\wedge \mathcal{E}  
        + \dualp{u} [\mathcal{B} ] \ ,
\eeq
with $\Tr \mathcal{A}^2 = -\mathcal{A}_{\alpha\beta} \mathcal{A}^{\alpha\beta}=2(\mathcal{E}^2 -\mathcal{B}^2)$ and 
$\det \mathcal{A} 
=-(\mathcal{E} \cdot \mathcal{B})^2$
and $\mathcal{E} \cdot \mathcal{B} = 
-%correction in proof
{\textstyle\frac14}\dual \mathcal{A}_{\alpha\beta} \mathcal{A}^{\alpha\beta}$.
The eigenvalue problem then finds invariant directions under this linear transformation
\beq
 \mathcal{A} X = \lambda X \rightarrow
  0=\lambda^4 - {\textstyle\frac12} \lambda^2 \Tr \mathcal{A}^2 + \det \mathcal{A}
  \ ,
\eeq
resulting in Synge's formula \cite{synge} for the eigenvalues $\lambda$
\beq
2 %correction in proof  
\lambda^2= (\mathcal{E}^2 -\mathcal{B}^2) \pm
        [(\mathcal{E}^2 -\mathcal{B}^2)^2+
4 %correction in proof
(\mathcal{E} \cdot \mathcal{B})^2]^{1/2}\ .
\eeq

One can easily classify the different cases which arise; using his terminology:
\begin{itemize}
\item general case: $\mathcal{E} \cdot \mathcal{B} \neq 0 $\\
     two real nonzero eigenvalues $\lambda=\pm \sigma_{(B)}$ summing to zero, each with a corresponding null eigenvector, two nonzero purely imaginary complex conjugate eigenvalues $\lambda=\pm i\sigma_{(R)}$ with a 2-dimensional spacelike eigenspace;
     generates a Lorentz 4-screw in a pair of orthogonal 2-planes, one timelike (boost plane), one spacelike (rotation plane).              

\item semi-singular case a):  $\mathcal{E} \cdot \mathcal{B} = 0 $
   and $\mathcal{B}^2 -\mathcal{E}^2>0$ (magnetic dominance)\\
   one pair of nonzero purely imaginary eigenvalues $\lambda=\pm i\sigma_{(R)}$ which sum to zero;
   one eigenvalue $\lambda=0$ with a 2-dimensional timelike eigenspace 
   which is pointwise invariant; 
   generates a rotation in an orthogonal spacelike 2-plane.

 \item semi-singular case b):   $\mathcal{E} \cdot \mathcal{B} = 0 $
  and $\mathcal{B}^2 -\mathcal{E}^2<0$ (electric dominance)\\
   two real nonzero eigenvalues $\lambda=\pm \sigma_{(B)}$ with two corresponding null eigenvectors and one $\lambda=0$ with a 2-dimensional spacelike eigenspace which is pointwise invariant;
  generates a boost in an orthogonal timelike 2-plane.                                                               

\item singular case: $\mathcal{E} \cdot \mathcal{B} = 0 $
   and $\mathcal{B}^2 -\mathcal{E}^2=0$ (null field)\\
   one eigenvalue $\lambda=0$ with a 2-dimensional null eigenspace which is pointwise invariant;
   generates a null rotation (also called a parabolic Lorentz transformation).

\end{itemize}

It is remarkable that one cannot easily find this discussion in a more recent textbook. In the general case one can always choose a new timelike unit vector $U$ with respect to which the electric and magnetic parts of $\mathcal{A}$ are aligned along the same direction, while in the semi-singular cases one can choose a new timelike unit vector such that either the magnetic or electric part vanishes respectively for the cases of electric and magnetic dominance.

\section*{Acknowledgments}
The authors gratefully acknowledge fruitful discussion with B. Mashhoon.


\section*{References}

\begin{thebibliography}{00}

\bibitem{rot}
Rothman T Ellis G F R and Murugan J 2001 
{\it Class.\ Quantum Grav.\/} {\bf 18} 1217

\bibitem{MMM}
Maartens R Mashhoon B and Matravers D R 2002
{\it Class.\ Quantum Grav.\/} {\bf 19} 195 

\bibitem{bjm}
Bini D Jantzen and R T Mashhoon B 2001
{\it Class.\ Quantum Grav.\/} {\bf 18} 1

\bibitem{mistau}
Misner C W and Taub A H 1968
{\it Zh.\ Eksp.\ Teor.\ Fiz.} {\bf 55} 233 [English translation:
{Sov.\ Phys. JETP\/} {\bf 55} 122 1969]

\bibitem{Miller}
Miller J G 1973
{\it J. Math.\ Phys.\/} {\bf 14} 486

\bibitem{Carter}
Carter B 1968
{\it Comm.\ Math.\ Phys.\/} {\bf 10} 280

\bibitem{DemNew}
Demia\'nski M and Newman E T 1966
{\it Bulletin de l'Acad\'emie Polonaise des Sciences} {\bf XIV} 653

\bibitem{zimsha}
Zimmerman R L and Shahir B Y 1989 
{\it Gen.\ Rel.\ Grav.\/} {\bf 21} 821

\bibitem{lynnou}
Lynden-Bell D and Nouri-Zonoz M 1998
{\it Rev.\ Mod.\ Phys.\/} {\bf 70} 427

\bibitem{noulyn}
Nouri-Zonoz M and Lynden-Bell D 1997
{\it Mon.\ Not.\ R.\ Astron.\ Soc.\/} {\bf 292} 714

\bibitem{synge} 
Synge J L 1965 
{\it Relativity: The Special Theory\/} 2nd Edition 
North Holland Amsterdam,  Chapter IV


\bibitem{bogiati}
Bollini C Giambiagi J J and Tiomno J 1981
{\it Nuovo Cimento Lett.} {\bf 31} 13 

\bibitem{idcf1}
Bini D Jantzen R T and Carini P 1997
{\it Int.\ J.\ Mod.\ Phys.\/} {\bf D6} 1

\bibitem{circfs} 
Bini D Merloni A and Jantzen R T  1999
{\it Class.\ Quantum Grav.\/} {\bf 16} 1333

\bibitem{mfg}
Jantzen R T, Carini P and Bini D 1992
  {\it Ann.\ Phys.\ (N.Y.)\/} {\bf 215} 1

\bibitem{kramer}
Kramer D  Stephani H  Herlt E and  MacCallum M A H,
{\it Exact Solutions of Einstein's Theory},
edited by E. Schmutzer, (Cambridge University Press, Cambridge, 1980)

\bibitem{mtw}
Misner C W Thorne K S and Wheeler J A 1973
{\it Gravitation} 
Freeman San Francisco  p.~481

\bibitem{idcf2}
Bini D Jantzen R T and Carini P 1997
{\it Int.\ J.\ Mod.\ Phys.\/} {\bf D6} 143
\end{thebibliography}
\end{document}